\documentclass[12pt]{iopart}
\usepackage{epsfig,rotate}

\def\re{\mathop{\rm Re}\nolimits}
\def\im{\mathop{\rm Im}\nolimits}
\begin{document}
\jl{1}
\title[Summability of the perturbative expansion for a
       disordered spin model]{Summability of the
       perturbative expansion for a
       zero-dimensional disordered spin model}
\author{Gabriel \'Alvarez\dag,
        Victor Mart\'{\i}n-Mayor\ddag\ and\\
        Juan J. Ruiz-Lorenzo\dag}
\address{\dag Departamento de F\'{\i}sica Te\'orica,
              Facultad de Ciencias F\'{\i}sicas,
	      Universidad Complutense, 28040 Madrid, Spain}
\address{\ddag Dipartimento di Fisica and INFN,
               Universit\`a di Roma ``La Sapienza'',
	       P.le Aldo Moro 2,
	       00185 Rome, Italy}
\begin{abstract}
We show analytically that the perturbative expansion for the free
energy of the zero dimensional (quenched) disordered Ising model
is Borel-summable in a certain range of parameters, provided that
the summation is carried out in two steps: first, in the strength
of the original coupling of the Ising model and subsequently in
the variance of the quenched disorder.  This result is illustrated
by some high-precision calculations of the free energy obtained by
a straightforward numerical implementation of our sequential
summation method.
\end{abstract}
\maketitle
\section{Introduction\label{sec:intro}}
One of the simplest cases to study the effects of impurities in
magnetic systems occurs in some doped uniaxial antiferromagnets such
as ${\rm Fe}_{0.46}{\rm Zn}_{0.54}{\rm F}_2$~\cite{SB99}, where the
${\rm Zn}$ impurities do not induce competing interactions and
Monte Carlo simulations with the three-dimensional random-site
Ising model reproduce accurately the behavior close to
the Curie temperature~\cite{BF98}.  Nevertheless, the successful
application of renormalization group techniques~\cite{ZJFT} to the
calculation of critical exponents for the pure Ising model makes
perturbative expansions a very appealing alternative to the Monte
Carlo calculations mentioned above.

The appropriate field theory for this problem is the usual
$u\phi^4$ theory in three dimensions modified with a random
mass term, and the analytic difficulties raised by the quenched
impurities are usually bypassed with the replica trick:  the system
is temporarily replaced by $n$ non-interacting copies for which the
disorder is now annealed, yielding an $O(n)$ theory with coupling
constant proportional to minus the variance of the quenched
disorder $w$, with an additional cubic anisotropy term with
coupling constant $u$.  This theory can be studied perturbatively
and the results corresponding to the original system recovered
(in principle) in the limit $n\rightarrow 0$.

Indeed, there are rather accurate analytic calculations of the
critical exponents of the random-site Ising model~\cite{FI3D}, but the
situation is less satisfactory than for pure systems~\cite{ZJFT}.  For
instance, the asymptotic parameter in disordered systems is
$\sqrt{\epsilon}$ instead of $\epsilon$ (where $\epsilon=4-d$, $d$
being the dimensionality of the space), and the $\beta$-function
computed at three loops shows no stable fixed point~\cite{FH99}.  In
an attempt to understand disordered systems in a simpler setting,
Bray~\etal\cite{BM87} and McKane~\cite{MC94} studied the asymptotic
expansion for the free energy in the zero-dimensional case, with
discouraging results: the toy-model was found non Borel-summable, in
sharp contrast with the (ordered) zero-dimensional $u\phi^4$ theory,
where Borel-summability has been rigorously
proved~\cite{ZJFT}. (Indeed, Borel summability in three dimensions
has been also proved~\cite{3DIM}.)

We consider in this paper the same zero-dimensional problem
of references~\cite{BM87,MC94}, and show that a slightly more
elaborated procedure can recover the free energy from the
perturbative expansion:  the Borel-summation has to be carried
out first in the strength of the original coupling of the Ising
model $u$ and subsequently in the variance of the quenched
disorder $w$.

The layout of the rest of the paper is as follows.  We first review
the main results of Bray~\etal and McKane in section~\ref{sec:pre};
the derivation of the Borel-summable double asymptotic expansion
for the free energy is carried out in the third section;
in section~\ref{sec:num} we give the details of our numerical
procedure and some illustrative examples, and the paper ends with
a brief summary.
\section{Previous Results\label{sec:pre}}

The free energy of the zero-dimensional Ising model  with quenched
dilution can be studied with a  $u\phi^4$ theory with a random mass
term
\begin{equation}
 f(u,w) = - \int_{-\infty}^{\infty} \frac{\rmd \psi}{\sqrt{4\pi w}}
            \,\e^{-\psi^2/4w} \log Z(\psi,u)
\end{equation}
where
\begin{equation}
 Z(\psi,u) = \int_{-\infty}^{\infty} \frac{\rmd \phi}{\sqrt{2\pi}}
             \,\e^{-\case12(1+\psi)\phi^2 - \case14 u\phi^4}
 \label{eq:z}
\end{equation}
is the partition function. Bray~\etal~\cite{BM87} fix
\begin{equation}
 \lambda = w/u > 0
\end{equation}
and write the asymptotic expansion for the free energy as
\begin{equation}
 f(u,w) \sim - \sum_{K=1}^\infty A_{K}(\lambda) u^{K}
 \label{eq:fak}
\end{equation}
where the $A_{K}(\lambda)$ are polynomials in $\lambda$ with
rational coefficients and degree $K$.  Then they use the
replica trick an a saddle point argument to infer that
the asymptotic behavior of $A_{K}(\lambda)$ as
$K\rightarrow\infty$ is given by two terms:
\begin{equation}
  A_{K}(\lambda) \sim
  \left\{
  \begin{array}{ll}
    A_{K}^{(1)}(\lambda) + A_{K}^{(\infty)}(\lambda), &
    0<\lambda<1 \\
    A_{K}^{(\infty)}(\lambda), & 1\leq\lambda
  \end{array}
  \right.
  \label{eq:akasy}
\end{equation}
The term $A_{K}^{(1)}(\lambda)$ is dominant for
$0<\lambda\le\case12$ and Borel-summable, while
\begin{equation}
 A_{K}^{\infty}(\lambda) = 
 - \frac{K!(4\lambda)^k}{\sqrt{\pi}K^{3/2}}
   \exp(-\gamma\sqrt{K}+\sigma)\cos(\mu\sqrt{K}+\delta)
\end{equation}
with coefficients $\gamma$, $\sigma$, $\mu$ and $\delta$
that they ultimately adjust by comparison between this
asymptotic formula and the numerical values of the
$A_{K}(\lambda)$ for $K$ up to 200 and several values
of $\lambda$ greater than $\case12$.  The increasing-period
cosine oscillations suggest them that there is an essential
singularity of the Borel-transform in the positive real $u$ axis,
and therefore the series is not Borel-summable for any
finite disorder (although the term $A_{K}^{(\infty)}(\lambda)$
is subdominant in the region $0<\lambda\le\case12$).
Despite the non-summability of the series~(\ref{eq:fak}),
i.e.~acknowledging that there would be some error even if they
could compute with all the terms, Bray~\etal set $u=1$ and
try to extract an accurate answer from~(\ref{eq:fak}) using
a conformal transformation suitable for oscillatory series,
but they find that as soon as $\lambda>\case12$ the method fails.

McKane~\cite{MC94} avoids the replica trick in favor of a more
direct analysis that clearly identifies the two sources of
non-analytic behavior of the free energy leading to the
two contributions is equation~(\ref{eq:akasy}): the branch cut
along the negative $u$ axis and the zeros of the partition
function~(\ref{eq:z}).  Moreover, he obtains the exact values of
$\gamma$, $\sigma$, $\mu$ and $\delta$ in terms of the zeros of
the partition function and guesses the form of the nonperturbative
contributions to the free energy that yield the asymptotic
behavior~(\ref{eq:akasy}).

The possibility of alternative summation methods, however,
remained unaddressed in these works.
\section{Summable asymptotic expansion for the free energy
         \label{sec:asy}}
For analyticity considerations, we remark that in terms of the new
variable
\begin{equation}
 z = \frac{1+\psi}{2 u^{1/2}}
\end{equation}
the partition function~(\ref{eq:z}) can be written as
\begin{equation}
 Z(\psi,u) = \frac{2^{1/2}}{\pi^{1/2} u^{1/4}}
	     \int_{0}^{\infty}
             \rmd t\,\e^{- z t^2 - \case14 t^4}
 \label{eq:zz}
\end{equation}
which in turn~\cite{MC94} can be expressed in terms of a parabolic
cylinder function or a modified Bessel function:
\begin{eqnarray}
 Z(\psi,u) & = & u^{-1/4} (2\pi)^{-1/2}
		 \e^{\case12 z^2} z^{1/2} K_{1/4}(\case12 z^2)
		 \label{eq:zk}\\
	   & = & (2u)^{-1/4} \e^{\case12 z^2}
	         D_{-1/2}(\sqrt{2} z).
		 \label{eq:zd}
\end{eqnarray}
There is, however, another expression particularly suitable for
later summability considerations: via a Mellin-Barnes
integral~\cite{SL60}, the partition function can be written as a
Laplace transform in $z^2$,
\begin{eqnarray}
 Z(\psi,u) & = & \frac{1}{(4u)^{1/4}\Gamma(\case14)}
             \int_{0}^{\infty}
	     \frac{\rmd t\,\e^{-z^2 t}}{[t(t+1)]^{3/4}}
             \qquad (\re (z^2) > 0)
 \label{eq:z2lt}\\
           & = & \frac{z}{(4u)^{1/4}\Gamma(\case14)}
             \int_{0}^{\infty}
	     \frac{\rmd t\,\e^{- t}}{[t(t+z^2)]^{3/4}}
             \qquad (\re z > 0).
 \label{eq:zlt}
\end{eqnarray}
Equations~(\ref{eq:zz}) or~(\ref{eq:zd}) show that
\begin{equation}
  F(z) = u^{1/4} Z(\psi,u)
 \label{eq:fdef}
\end{equation}
is an entire function of $z$, and from equation~(\ref{eq:zk})
and reference~\cite{WA44} it follows that the only zeros
of $F(z)$ appear in complex-conjugate pairs $z_{k}, \bar{z}_{k}$
in the $\re z<0$ half-plane, and are given asymptotically by
\begin{equation}
 z_{k}^2 \sim - \case12\log 2 - \rmi\case\pi{2} (4k+3)
 \qquad (\im z_{k}>0, k = 0,1,2,\ldots).
\end{equation}
To stay close to the notation of Bray \etal and McKane and yet
show that we are dealing with a two-variable problem, we set
\begin{equation}
 \lambda = w/u
\end{equation}
\begin{equation}
 \mu = 1/(2 u^{1/2})
\end{equation}
\begin{equation}
 \Phi(\mu,\lambda)  = f(u,w).
\end{equation}
In terms of these variables, the free energy can be written as
\begin{equation}
 - \Phi(\mu, \lambda) = \case12\log(2\mu) +
             \frac{1}{(\lambda\pi)^{1/2}}
	     \int_{-\infty}^{\infty}
	     \rmd z\, \e^{- (z - \mu)^2/\lambda}
	     \log[F(z)]
 \label{eq:phi}
\end{equation}
were the integral in the right-hand side is taken along the real
axis, and is well defined for any complex $\mu$ and $\re\lambda>0$.
Note that as $z$ varies along the real axis in the increasing sense,
$F(z)$ traces the positive axis in the decreasing sense, but as
we shift (say) upwards the integration path and $z$ picks up a
constant (positive) imaginary part
[i.e. $z=t+i\epsilon, t\in(-\infty,\infty)$], equation~(\ref{eq:zd})
shows that $F(z)$ traces a spiral ending at the origin for $t=\infty$.
In particular, if $z=t+z_{k}$ the spiral passes again through the
origin (see figure~1).  Therefore, if we shift vertically the
integration path to $z=t+\mu$ (with $0<\im\mu\neq\im z_{k}$),
we pick up a finite number of  contributions from the zeros of
$F(z)$ whose evaluation in terms of the complementary error function
is given by the following self-explanatory chain of equations:
\begin{eqnarray}
  \fl
  - \Phi(\mu, \lambda)  =  \case12 \log(2\mu) +
             \frac{1}{(\lambda\pi)^{1/2}}
	     \int_{-\infty+\mu}^{\infty+\mu}
	     \rmd z\, \e^{- (z - \mu)^2/\lambda}
	     \log[F(z)]\nonumber\\
 {}+
	     \sum_{0<\im z_{k}<\im\mu}
             \frac{1}{(\lambda\pi)^{1/2}}
	     \int_{-\infty,z_{k}^{(+)}}
	     \rmd z\, \e^{- (z - \mu)^2/\lambda}
	     \log[F(z)]\\
  \lo= \case12 \log(2\mu) +
      \frac{1}{(\lambda\pi)^{1/2}}
	     \int_{-\infty}^{\infty}
	     \rmd t\, \e^{- t^2/\lambda}
	     \log[F(t+\mu)]\nonumber\\
 {}+
	     \sum_{0<\im z_{k}<\im\mu}
             \frac{2\pi\rmi}{(\lambda\pi)^{1/2}}
	     \int_0^\infty
	     \rmd t\, \e^{- (z_k - t - \mu)^2/\lambda}\\
  \lo= \case12 \log(2\mu) +
      \frac{1}{(\lambda\pi)^{1/2}}
	     \int_{-\infty}^{\infty}
	     \rmd t\, \e^{- t^2/\lambda}
	     \log[F(t+\mu)]\nonumber\\
 {}+
	     \rmi\pi\sum_{0<\im z_{k}<\im\mu}
      {\rm erfc}\left(\frac{\mu-z_{k}}{\lambda^{1/2}}\right).
   \label{eq:erfc}
\end{eqnarray}

\begin{figure}[t!]
\begin{center}
\leavevmode
\epsfig{file=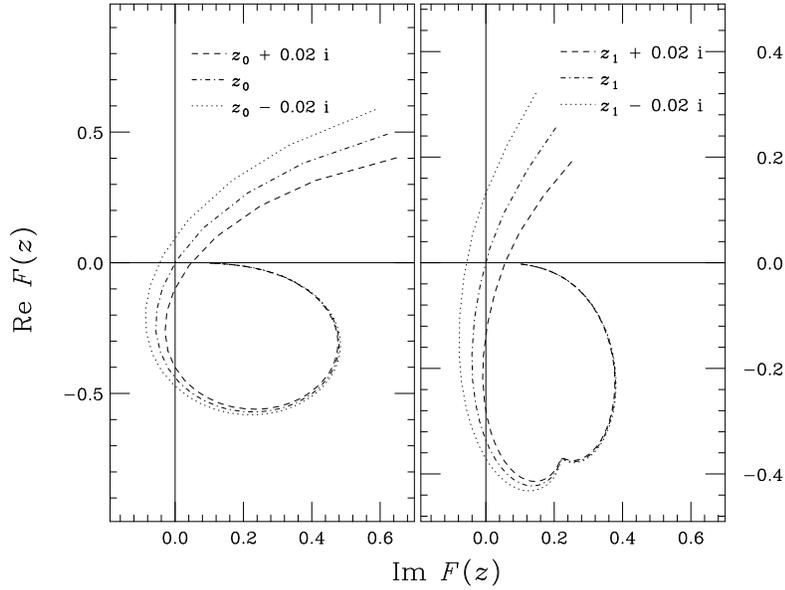,width=0.5\linewidth,angle=90}
\end{center}
\caption{Paths in the complex plane followed by the function $F(z)$
        as $z$ varies along straight lines parallel to the real
        axis slightly below, on and slightly above the first two
        zeros of $F(z)$ in the upper half plane
        ($z=z_k+t$, $z=z_k\pm {\mathrm i}\epsilon +t$,
        $-\infty<t<\infty$). The dip in the right plot is a trace
        of the first loop of the spiral (shown in the left plot)
        which unfolds via a cusp (when $F'(z)=0$) as the integration
        path shifts upwards.}
\label{FIG1}
\end{figure}

There are three key points to note in these expressions.
First, the determination of the logarithms: they are real for 
real argument (as $t\rightarrow\infty$) and must be followed
by continuity according to figure~1.
Second, for $|\im\mu|<\im z_0$ there are no contributions from
the zeros and we do have
\begin{equation}
 - \Phi(\mu, \lambda) = \case12\log(2\mu) +
             \frac{1}{(\lambda\pi)^{1/2}}
	     \int_{-\infty}^{\infty}
	     \rmd z\, \e^{-z^2/\lambda}
	     \log[F(z+\mu)].
 \label{eq:fzm}
\end{equation}
\begin{table}
\caption{Comparison among the $K=3$ partial sum of
equation~(\ref{eq:fak}), the shifted integral~(\ref{eq:fzm}),
the shifted integral plus corrections~(\ref{eq:erfc}), and the
unshifted integral~(\ref{eq:phi}) for $\lambda=1$ and three
values of $\mu$ with $0\leq\im\mu<\im z_{0}$,
$\im z_{0}<\im\mu<\im z_{1}$ and $\im z_{1}<\im\mu<\im z_{2}$
respectively.}

\begin{tabular}{lccc}
\br
            & $\mu=2$ & $\mu=1+2\rmi$ & $\mu=1+5\rmi/2$ \\
\mr
Partial sum & $0.0155640$
            & $-0.00752925-\rmi 0.0099890$
	    & $-0.00624927-\rmi 0.00593754$ \\
Equation~(\ref{eq:fzm}) & $0.0155379$
                        & $-0.00752689-\rmi 0.0099516$
			& $-0.00624739-\rmi 0.00593804$ \\
Equation~(\ref{eq:erfc})
            & $0.0155379$
            & $-0.00620929-\rmi 0.0107989$
	    & $-0.00908002-\rmi 0.00573683$ \\
Equation~(\ref{eq:phi})
            & $0.0155379$
            & $-0.00620929-\rmi 0.0107989$
	    & $-0.00908002-\rmi 0.00573683$ \\
\br
\end{tabular}
\end{table}
We stress that the right-hand side of equation~(\ref{eq:fzm})
is well defined whenever $\im\mu\neq\im z_{k}$, but it is equal
to the free energy only for $|\im\mu|<\im z_0$.  As a numerical
illustration of this point, in the last three rows of table~1
we show numerical calculations of the shifted integral~(\ref{eq:fzm}),
the shifted integral plus corrections~(\ref{eq:erfc}), and the
unshifted integral~(\ref{eq:phi}) for $\lambda=1$ and three
values of $\mu$ with $0\leq\im\mu<\im z_{0}$,
$\im z_{0}<\im\mu<\im z_{1}$ and $\im z_{1}<\im\mu<\im z_{2}$
respectively.
Third, equation~(\ref{eq:fzm}) can be rewritten as a Laplace
transform in $\lambda^{-1}$
\begin{equation}
 - \Phi(\mu,\lambda) = \case12\log(2\mu) +
    \frac{1}{\lambda^{1/2}}
    \int_{0}^{\infty} \rmd t\,\e^{-t/\lambda} B(\mu,t)
 \label{eq:bmt}
\end{equation}
where
\begin{equation}
 B(\mu, t) = \frac{1}{(4\pi t)^{1/2}}
        \log\left[F(\mu+t^{1/2}) F(\mu-t^{1/2})\right].
 \label{eq:bdef}
\end{equation}
The singularities of $B(\mu, t)$ as a function of $t$ are readily
calculated in terms of the $z_{k}$ and (apart from the trivial
singularity at the origin) stay at a finite distance of the
positive real axis whenever $\im\mu\neq\im z_{k}$.

Equations~(\ref{eq:zlt}) and~(\ref{eq:fdef}) imply that we can
obtain a Borel-summable asymptotic expansion for $F(z)$ and its
derivatives, and equations~(\ref{eq:bmt}) and~(\ref{eq:bdef})
that, as far as we stay away from the zeros of $F$, we can also
obtain a Borel-summable series in $\lambda$.
Our final result will be, therefore, a Borel-summable series
(in $\lambda$) with each of its coefficients given
by a Borel-summable series (in $\mu^{-2}=4u$). 

Now we proceed to the details of the derivation. 
We first expand $\log[F(z+\mu)]$ in equation~(\ref{eq:fzm})
in (convergent) Taylor series around $z=0$---or equivalently $B(\mu,t)$ in
equation~(\ref{eq:bmt}) around $t=0$---and integrate term by term to get
\begin{equation}
 - \Phi(\mu, \lambda) \sim \case12\log(2\mu) +
           \sum_{n=0}^{\infty}
	   (\log\circ F)^{(2n)}(\mu) \frac{\lambda^n}{4^{n}n!}.
\end{equation}
Then, we use the Borel-summable asymptotic expansion for $F(z)$
that follows directly from equation~(\ref{eq:zlt})
\begin{equation}
 F(z) \sim (2 z)^{-1/2} {}_{2}F_{0}(\case14,\case34;;-z^{-2})
      \qquad (-\case\pi{2}< \arg z <\case\pi{2})
\end{equation}
to obtain asymptotic expansions for the derivatives
$(\log\circ F)^{(2n)}(\mu)$.
Incidentally, we mention that the summability sector of $F(z)$
in terms of $u$ is just $-\pi<\arg u<\pi$, i.e. we are rederiving
the well-known summability of the (ordered) zero-dimensional $u\phi^4$
theory as explained, for example, in reference~\cite{ZJFT}.

Note that for the zero-th derivative we have
\begin{eqnarray}
 \log[F(\mu)] & \sim & - \case12\log(2 \mu) +
      \log\left[{}_{2}F_{0}(\case14,\case34;;-\mu^{-2})\right]\\
             & = & - \case12\log(2 \mu) +
	           \sum_{k=1}^{\infty} \frac{(-1)^k b_{k}}{\mu^{2k}}
 \label{eq:logfmu} 
\end{eqnarray}
and the logarithm cancels the first term in the right-hand
side of equation~(\ref{eq:fzm}). (Equation~(\ref{eq:logfmu}) is in fact
the definition of the coefficients $b_{k}$).  Higher derivatives have
pure asymptotic power series that can be computed easily in terms of
the zero-th derivative series
\begin{equation}
 (\log\circ F)^{(2n)}(\mu) \sim \frac{1}{\mu^{2n}}
 \left[     
       \frac{\Gamma(2n)}{4} +
       \sum_{k=1}^{\infty}
       \frac{\Gamma(2n+2k)}{\Gamma(2n)}\frac{(-1)^k b_{k}}{\mu^{2k}}
 \right].
 \label{eq:f2n}
\end{equation}
The final result is the double series
\begin{equation}
 \fl
 - \Phi(\mu, \lambda) \sim
   \sum_{k=1}^{\infty} \frac{(-1)^k b_{k}}{\mu^{2k}} +
   \sum_{n=1}^{\infty}
   \frac{1}{n!}
   \left(\frac{\lambda}{4 \mu^2}\right)^n
   \left[     
       \frac{\Gamma(2n)}{4} +
       \sum_{k=1}^{\infty}
       \frac{\Gamma(2n+2k)}{\Gamma(2n)}\frac{(-1)^k b_{k}}{\mu^{2k}}
   \right]
 \label{eq:dsf}
\end{equation}
whose structure in terms of the original variables is ($c_{00}=0$)
\begin{equation}
 - f(u, w) \sim \sum_{n=0}^{\infty}
                \left(
		      \sum_{k=0}^{\infty} c_{n,k} u^k
		\right) w^n\, .
\end{equation}
This double series is formally equivalent to equations~(B10),
(B11) and~(B13) in reference~\cite{BM87} (the only difference
is that we have transposed the indices in $c_{n,k}$),
but with a definite ordering that cannot be altered
without further considerations:  to take advantage of the
summability we should first Borel-sum the series in $u$
(see the precise definition in the next section)
\begin{equation}
    c_{n}(u) = \mbox{Borel}\sum_{k=0}^{\infty} c_{n,k} u^{k}
\end{equation}
and then in $w$
\begin{equation}
    -f(u,w) = \mbox{Borel}\sum_{n=0}^{\infty} c_{n}(u) w^n
\end{equation}
instead of performing first the finite sum
\begin{equation}
 A_{K}(\lambda) = \sum_{n=0}^{K}c_{n,K-n}\lambda^n
\end{equation}
followed by some kind of summation in $u$.  Although this last
reordering preserves the asymptotic nature of the series
to the function defined by equation~(\ref{eq:fzm}) (as we also
illustrate in the first row of table~1), the numerical and
analytic evidence presented by Bray~\etal  shows that
it spoils the summability of equation~(\ref{eq:dsf}).
\section{Numerical algorithm and results\label{sec:num}}
Of the two standard methods to achieve the analytic continuation
implicit in the Borel-summation (conformal mapping or Pad\'e
approximants) we have chosen the second because it does not
require a precise knowledge of the singularities of the function,
although some Pad\'e approximants may introduce spurious
singularities close to the integration path.  Equation~(\ref{eq:f2n})
suggests that an appropriate numerical Borel-sum of the $u$ series is
\begin{equation}
    c_{n}(u) \approx \frac{1}{u^{2n+1/2}}
    \int_{0}^{\infty} \e^{-t/u} t^{2n-1/2} P^{[p,q]}(t) \,\rmd t
    \label{eq:nbt}
\end{equation}
where $ P^{[p,q]}(t)$ is the $[p,q]$-Pad\'e approximant for
\begin{equation}
    \hat{c}_{n}(t) = \sum_{k=0}^{p+q}
                     \frac{c_{n,k} t^{k}}{\Gamma(2n+k+\case12)}.
\end{equation}
Equation~(\ref{eq:nbt}) has the additional advantage over other
equivalent forms that the Pad\'e approximants and their zeros
(which we use later) need to be calculated only once.  Furthermore,
to avoid problems with the numerical integration we expand the
Pad\'e approximant as a polynomial plus partial fractions, i.e.
assuming that all the poles are simple
\begin{equation}
    P^{[p,q]}(t) = \sum_{k=0}^{p-q} p_{k} t^{k} +
                   \frac{R(t)}{S(t)}
	         = \sum_{k=0}^{p-q} p_{k} t^{k} +
		   \sum_{k=1}^{q}
	           \frac{R(t_{k})}{S'(t_{k})(t-t_{k})}
\end{equation}
and the integration in equation~(\ref{eq:nbt}) can be carried out
in terms of complete and incomplete gamma functions evaluated at
the $q$ poles $t_{k}$ of $P^{[p,q]}(t)$:
\begin{eqnarray}
    \fl
    c_{n}(u) \approx
             \sum_{k=0}^{p-q} p_{k} u^k \Gamma(2n+k+\case12)
	     \nonumber\\
	     {}+
	     \frac{\Gamma(2n+\case12)}{u^{2n+1/2}}
             \sum_{k=1}^{q} \frac{R(t_{k})}{S'(t_{k})}
	     \e^{-t_{k}/u} (-t_{k})^{2n-1/2}
	     \Gamma(-2n+\case12,-t_{k}/u).
\end{eqnarray}
We have performed our calculations with the same 200 terms of
the expansion~(\ref{eq:logfmu}) used in reference~\cite{BM87}, using 
the arbitrary precision numerical capabilities of  {\sl Mathematica}.
Figure~2 shows some typical results of this first stage of our
double-summation method.  For four values of $u$ and the lowest
eleven coefficients $c_{0}(u),\ldots,c_{10}(u)$ we plot a measure
$\Delta$ of the number of correct digits to the right of the
decimal point,
\begin{equation}
 \Delta = -\log_{10}|c_{n}(u)^{\mbox{exact}}-
                     c_{n}(u)^{\mbox{summed}}|
\end{equation}
as a function of the order $p$ of the diagonal Pad\'e
approximants $P^{[p,p]}(t)$ in equation~(\ref{eq:nbt}).
As a general trend, the precision of the summed coefficients
increases with the order of the approximant, although with smaller
slope for higher values of $u$ (note that we have used the same
scales in the four plots).  The irregularities in figure~2 are
often due to ill-conditioned linear systems of equations in the
calculation of the Pad\'e approximants, and are usually corrected
using higher precision or higher order approximants.  We have also
checked that neighboring off-diagonal Pad\'e approximants give
similar results.
\begin{figure}[t!]
\begin{center}
\leavevmode
\epsfig{file=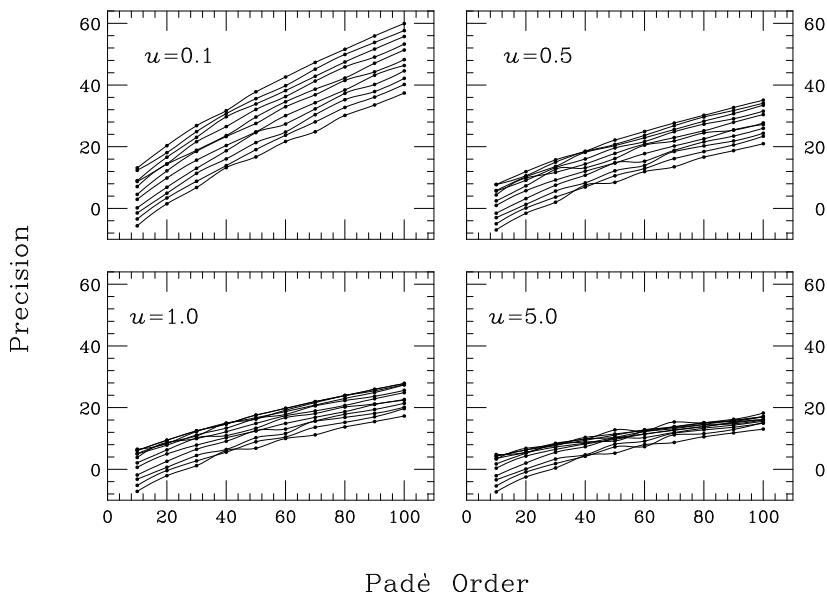,width=0.5\linewidth,angle=90}
\end{center}
\caption{Number of correct digits to the right of the decimal point
        in the Borel-summed coefficients $c_{0}(u),\ldots,c_{10}(u)$
        (from top to bottom at $p=10$) 
        as a function of the order of the diagonal Pad\'e
        approximant in equation~(\ref{eq:nbt}).}
\label{FIG2}
\end{figure}

The second stage, the summation in $w$, is equivalent to the
$n=0$ summation of the first stage.  Now the Borel-transformed
coefficients $c_n(u)/\Gamma(n+\case12)$ are precisely the coefficients
of the convergent Taylor expansion of $t^{1/2} B(\mu,t)$,
and to illustrate how well the Pad\'e approximants continue the
Taylor series beyond the radius of convergence, we plot
in figure~3 the exact integrand $B(\case12,t)$
corresponding to $u=1$ in equation~(\ref{eq:bmt}), and its
approximations $t^{-1/2} P^{[p,p]}(t)$ for $p = 10$, $15$ and $20$.
The radius of convergence of the Taylor series is marked in the plot
by the vertical line at $t = |z_0 - \case12|^{2} \approx 6.54$.
\begin{figure}[t!]
\begin{center}
\leavevmode
\epsfig{file=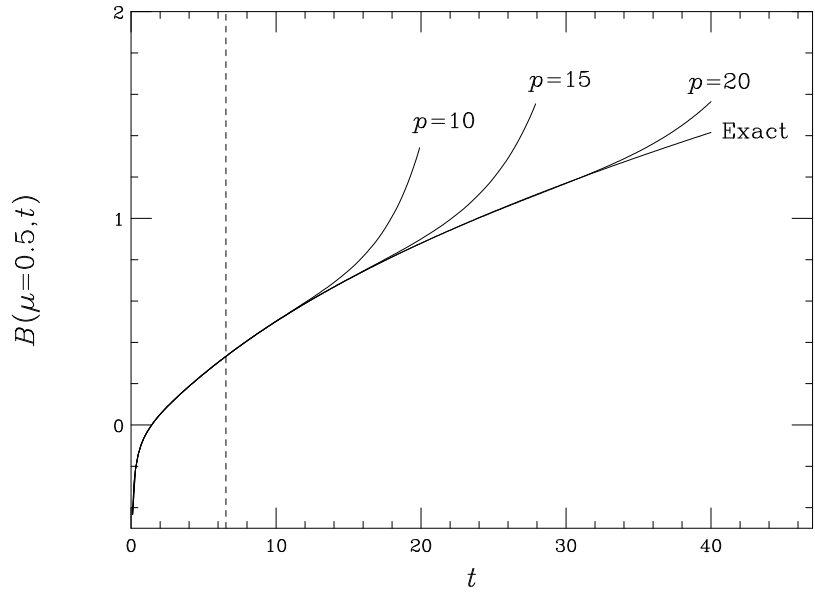,width=0.5\linewidth,angle=90}
\end{center}
\caption{Exact integrand $B(\case12,t)$ corresponding to $u=1$
        in equation~(\ref{eq:bmt}) and its approximations
        $t^{-1/2} P^{[p,p]}(t)$ for $p = 10, 15$ and $20$.
        The vertical line marks the radius of convergence of the
        Taylor series for $B(\case12,t)$.}
\label{FIG3}
\end{figure}

In figure~4 (again with $u=1$ and for several values of $\lambda$)
we plot the number of correct digits to the right
of the decimal point in the Borel-summed free energy as a function
of the order of the diagonal Pad\'e approximant.  We have performed
the calculations with all the $c_{n}(u)$ of the same precision,
so that the numerical error is due only to this second-stage summation.
The first feature worth mentioning is the steady increase in the
precision of the summed free energy (with a few exceptions that are
related again to the Pad\'e approximants), although the convergence
is slower for larger values of $\lambda$.  Note that the plateau in
the summation of Bray~\etal for $\lambda=0.2$ ends approximately at
fifty terms of the series, while at $p=40$ (i.e. with 81 terms of
the series) we have steady numerical convergence; $\lambda=0.5$
corresponds to the switch between asymptotic behaviors in
equation~(\ref{eq:akasy}), while in our reordering it is not
a distinguished point; finally, for $\lambda=0.8$ Bray~\etal are
already unable to get a good result, while our summation
still gives four correct digits for $\lambda=5$.

\begin{figure}[t!]
\begin{center}
\leavevmode
\epsfig{file=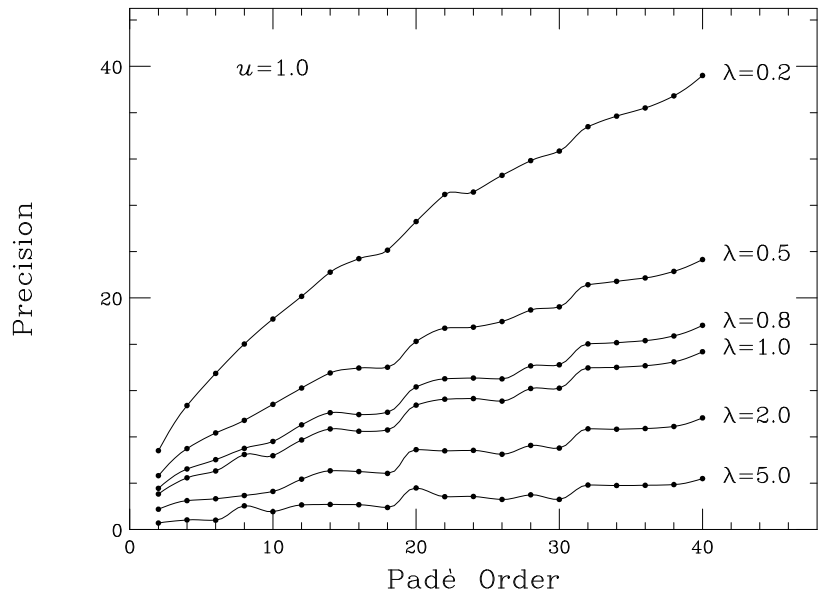,width=0.5\linewidth,angle=90}
\end{center}
\caption{Number of correct digits to the right of the decimal point
        in the Borel-summed free energy $f(1,\lambda)$ as a function
        of the order of the diagonal Pad\'e approximant used in the
        second stage of the double summation.}
\label{FIG4}
\end{figure}

\section{Summary\label{sec:sum}}
We have analyzed the Borel-summability of the perturbative expansion
for the free energy in one of the simplest disordered systems,
the zero dimensional quenched diluted Ising model.  This expansion
is a double asymptotic series in the original coupling $u$ of the
Ising model and in the variance $w$ of the quenched disorder,
and we have shown that the expansion is in fact Borel-summable by
a sequential method: first we sum in the coupling $u$ and
subsequently in the variance $w$.  This summability results
are valid at least for $-\pi<\arg u<\pi$ and $\re w>0$, with the
understanding that if we want to recover the free energy and
not the integral~(\ref{eq:fzm}), additional ${\rm erfc}$ terms are
needed in the domains specified in section~\ref{sec:asy}.

A straightforward numerical implementation of our double summation
method has permitted us to obtain (with the same number of terms of
the original series) accurate free energies for values of the
parameters well beyond the region where Bray~\etal were already
unable to get an estimate.

We expect that the insight provided by these zero dimensional
results will be helpful in the understanding of the more
complicated higher-dimensional case, despite the significant
qualitative differences motivated by Griffiths
singularities~\cite{GR69} in the generalized sense of
Dotsenko~\cite{DO99}.  In fact we feel that a clear
understanding of the Griffiths phase in the field-theoretical
setting will be needed before the problem will be finally solved.

Finally, we would like to mention as open problems the study
of the negative $\lambda$ region, and a more thorough understanding
of the solution of this problem using replicas.
\ack
We thank Professors~G~Parisi for useful discussions and A~J~Bray for
providing references.  The financial support of the Universidad
Complutense under project PR156/97-7100, the Comisi\'on
Interministerial de Ciencia y Tecnolog\'{\i}a under project
AEN96-1708, and a postdoctoral grant from the MEC are also gratefully
acknowledged.


\section*{References}
\begin{thebibliography}{00}
\bibitem{SB99} Slani\v{c} Z, Belanger D P and
               Fernandez-Baca J A 1999
               {\it Phys. Rev. Lett.} {\bf 82} 426
\bibitem{BF98} Ballesteros H G, Fernandez L A, Mart\'{\i}n-Mayor V,
               Mu\~noz Sudupe A, Parisi G and Ruiz-Lorenzo J J 1998
               {\it Phys. Rev. B} {\bf 58} 274
\bibitem{ZJFT} Zinn-Justin J 1990
               {\it Quantum Field Theory and Critical Phenomena}
		             (Oxford: Oxford Science Publications)
\bibitem{FI3D} Newman K E and Rieded E K 1982
               {\it Phys. Rev. B} {\bf 25} 264
\par\item[]    Jug G 1983
               {\it Phys. Rev. B} {\bf 27} 609
\par\item[]    Mayer I O 1989
               {\it J. Phys. A: Math. Gen.} {\bf 22} 2815
\bibitem{FH99} Folk R, Holovatch Y and Yavors'kii T 1999 
               {\it Pis'ma v ZhETF} {\bf 69} 698
               [{\it JETP Lett.} {\bf 69} 747] 
\bibitem{BM87} Bray A J, McCarthy T, Moore M A,
               Reger J D and Young A P 1987
               {\it Phys. Rev. B} {\bf 36} 2212
\bibitem{MC94} McKane A J 1994
               {\it Phys. Rev. B} {\bf 49} 12003
\bibitem{3DIM} Eckmann J P, Magnen J and S\'en\'eor R 1975
               {\it Comm. Math. Phys.} {\bf 39} 251 
\bibitem{SL60} Slater L J 1960
               {\it Complex Hypergeometric Functions}
		             (Cambridge: Cambridge University Press)
\bibitem{WA44} Watson G N 1944
               {\it A Treatise on the Theory of Bessel Functions}
		             (Cambridge: Cambridge University Press)
\bibitem{GR69} Griffiths R 1969
               {\it Phys. Rev. Lett.} {\bf 23} 17
\bibitem{DO99} Dotsenko V 1999
               {\it J. Phys. A: Math. Gen.} {\bf 32} 2949
\end{thebibliography}
\end{document}